\documentclass[journal,a4paper,onecolumn]{IEEEtran}

\usepackage{amsmath,amssymb}
\usepackage{graphicx}% Include figure files
\usepackage{dcolumn}% Align table columns on decimal point
\usepackage{bm}% bold math
\usepackage{siunitx}
\usepackage[utf8]{inputenc}
\usepackage[T1]{fontenc}
\usepackage{mathptmx}
\usepackage{etoolbox}

\makeatletter
\def\@email#1#2{%
 \endgroup
 \patchcmd{\titleblock@produce}
  {\frontmatter@RRAPformat}
  {\frontmatter@RRAPformat{\produce@RRAP{*#1\href{mailto:#2}{#2}}}\frontmatter@RRAPformat}
  {}{}
}%
\makeatother
\begin{document}
%\DeclareMathOperator{\sech}{sech}

%\preprint{AIgP/123-QED}

\title{Gate tunable lateral 2D \textit{pn} junctions: an analytical study of its electrostatics}
% Force line breaks with \\
\author{Ferney A. Chaves, Anibal Pacheco-Sanchez, David Jiménez

\thanks{This work has received funding from the European Union’s Horizon 2020 research and innovation programme under grant agreement No GrapheneCore3 881603, from Ministerio de Ciencia, Innovación y Universidades under grant agreements RTI2018-097876-B-C21(MCIU/AEI/FEDER, UE) and FJC2020-046213-I. This  article  has been partially  funded  by  the  European Union Regional Development Fund within the framework of the ERDF Operational Program of Catalonia 2014-2020 with the support of the Department de Recerca i Universitat, with a grant of 50\% of total cost eligible. GraphCAT project reference: 001-P-001702. We also acknowledge financial support by Spanish government under the project PID2021-127840NB-I00 (MCIN/AEI/FEDER, UE). \newline The authors are with the Departament d'Enginyeria Electr\`{o}nica, Escola d'Enginyeria, Universitat Aut\`{o}noma de Barcelona, Bellaterra 08193, Spain, e-mails: FerneyAlveiro.Chaves@uab.cat, AnibalUriel.Pacheco@uab.cat, David.Jimenez@uab.cat}
}
\maketitle
\makeatletter
\def\ps@IEEEtitlepagestyle{
  \def\@oddfoot{\mycopyrightnotice}
  \def\@evenfoot{}
}
\def\mycopyrightnotice{
  {\footnotesize
  \begin{minipage}{\textwidth}
  \centering
© 2023 IEEE. Personal use of this material is permitted. Permission from IEEE must be obtained for all other uses, in any current or future media, including reprinting/republishing this material for advertising or promotional purposes, creating new collective works, for resale or redistribution to servers or lists, or reuse of any copyrighted component of this work in other works.
  \end{minipage}
  }
}
\maketitle
\begin{abstract}
\boldmath
The electrostatics of two-dimensional (2D) lateral \textit{pn} homojunctions considering the impact of electrostatic doping by means of two split bottom-gates are studied here. Analytical expressions are obtained from the solution of the 2D Poisson equation considering a depletion approximation. Straightforward analytical models for the electrostatic potential and the depletion width within both the dielectric and the 2D semiconductor are obtained for both the symmetrical and asymmetrical cases. In contrast to the case of devices with chemical doping, the obtained depletion width model of devices with electrostatic doping do not depend on the dielectric constant but only on the electrostatic potential and oxide thickness. The models describe the electrostatics of gate-tunable 2D \textit{pn} junctions at arbitrary bias. A benchmark against numerical device simulations of MoS$_2$-based \textit{pn} junctions validate the analytical models.
\end{abstract}


\section{\label{sec:level1}Introduction}

Two-dimensional (2D)  based devices have been in the spotlight of the continuous downscaling trend of electronics due to their semiconductor thin-body, high-performance and integration feasibility into standard production lines \cite{AkiKuy19}. 2D \textit{pn} junctions are among the most versatile emerging devices for both electronics and optoelectronics applications \cite{WanPei21}. Homojunctions and heterojunctions have been demonstrated with 2D materials with various levels of reproducibility depending on the fabrication approach \cite{WanPei21,LiuHua19}. In contrast to chemical doping, electrostatic doping by a pair of bottom-gate contacts separated by a gap has been demonstrated to be a more straightforward solution in controlling the different carrier concentration regions in 2D lateral (L) PN junctions. Proof-of-concept devices based on these gate tunable (GT) 2D lateral \textit{pn} junctions have been experimentally proven to be suitable for optoelectronics \cite{PopFur14}-\cite{BieGro17}, electronics \cite{PanChe19} and neuromorphic applications \cite{PanChe19,PanWan20}. 

Modeling approaches have been presented elsewhere \cite{AchYes02}-\cite{ChaJim18} for chemically doped 2D \textit{pn} junctions, however, descriptions of the internal physical phenomena in GT 2D junctions are scarce due to the few theoretical studies on this structure \cite{ChaJim21}. In contrast to three-dimensional PN junctions where only a one-dimensional (1D) Poisson equation and the complete-depletion approximation are considered , electrostatics analysis in 2D junctions involves solving a 2D Poisson equation and the consideration of partial depletion in the transition between the depletion and quasi-neutral regions due to a weaker screening of the electric field \cite{NipJay17,ChaJim18}. Numerical device simulations and semi-analytical solutions have been proposed for GT junctions previously for the specific case of symmetrical gate voltages \cite{ChaJim21}. 

In this work, a general analysis of the electrostatics of GT lateral 2D \textit{pn} homojunctions, embracing both symmetric and asymmetric gate voltage tunning, yields a compact analytical model for the electrostatic potential at any location of the device cross-section. This work is organized as follows. The analytical models for the electrostatic potential and depletion width of the device under study are presented in section II. In section III, the models are applied to MoS$_2$ lateral pn junctions under different bias and with different device geometry and the results are discussed benchmarked with the output of an in-house numerical device simulation (NDS) tool \cite{ChaJim21}. A conclusion is provided in section IV followed by three useful appendix with details on the procedures to solve and evaluate equations in the main text.

\section{Device electrostatics}

\begin{figure}[htb!]
 \includegraphics[width=0.45\textwidth]{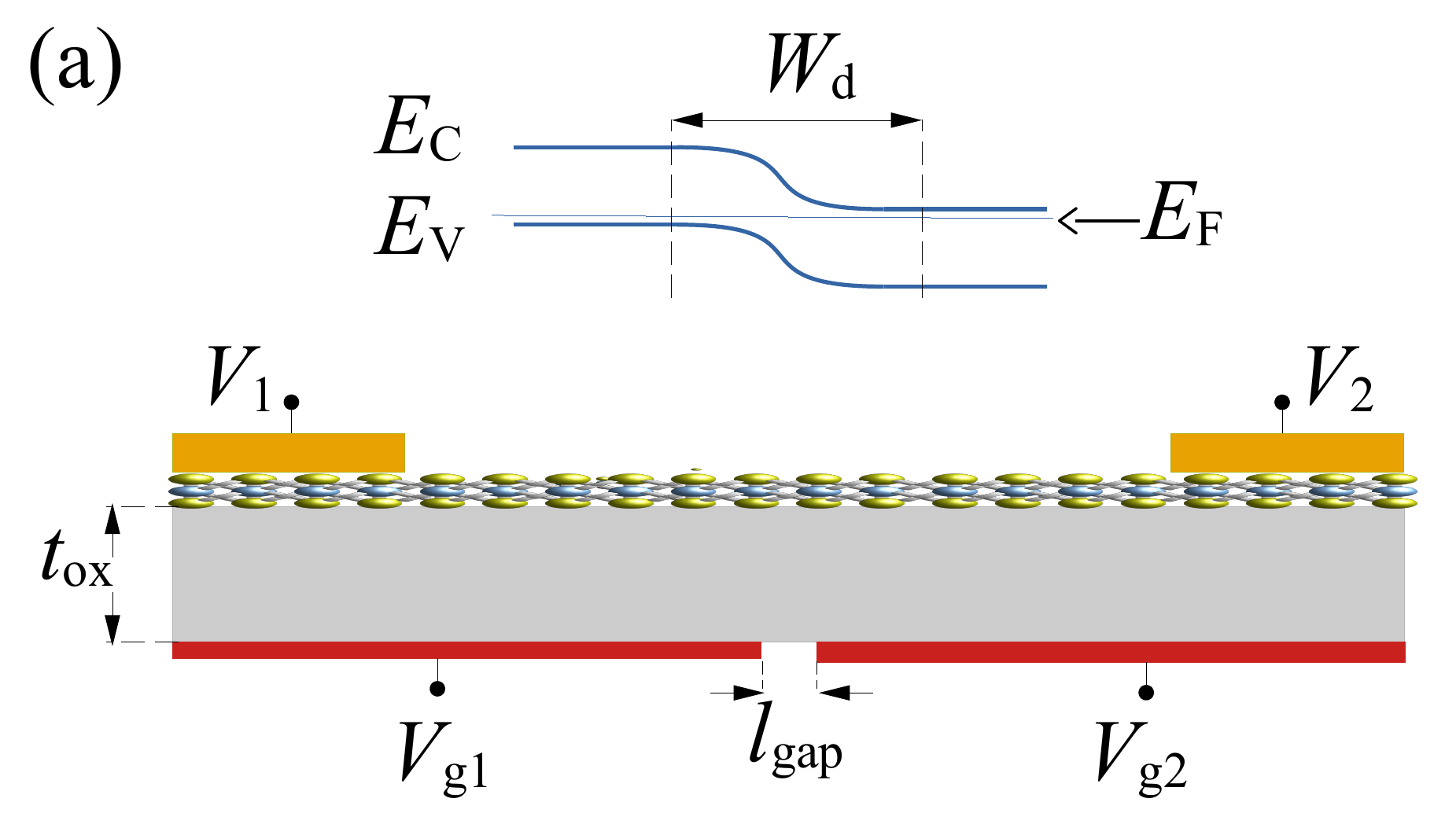}
  \includegraphics[width=0.425\textwidth]{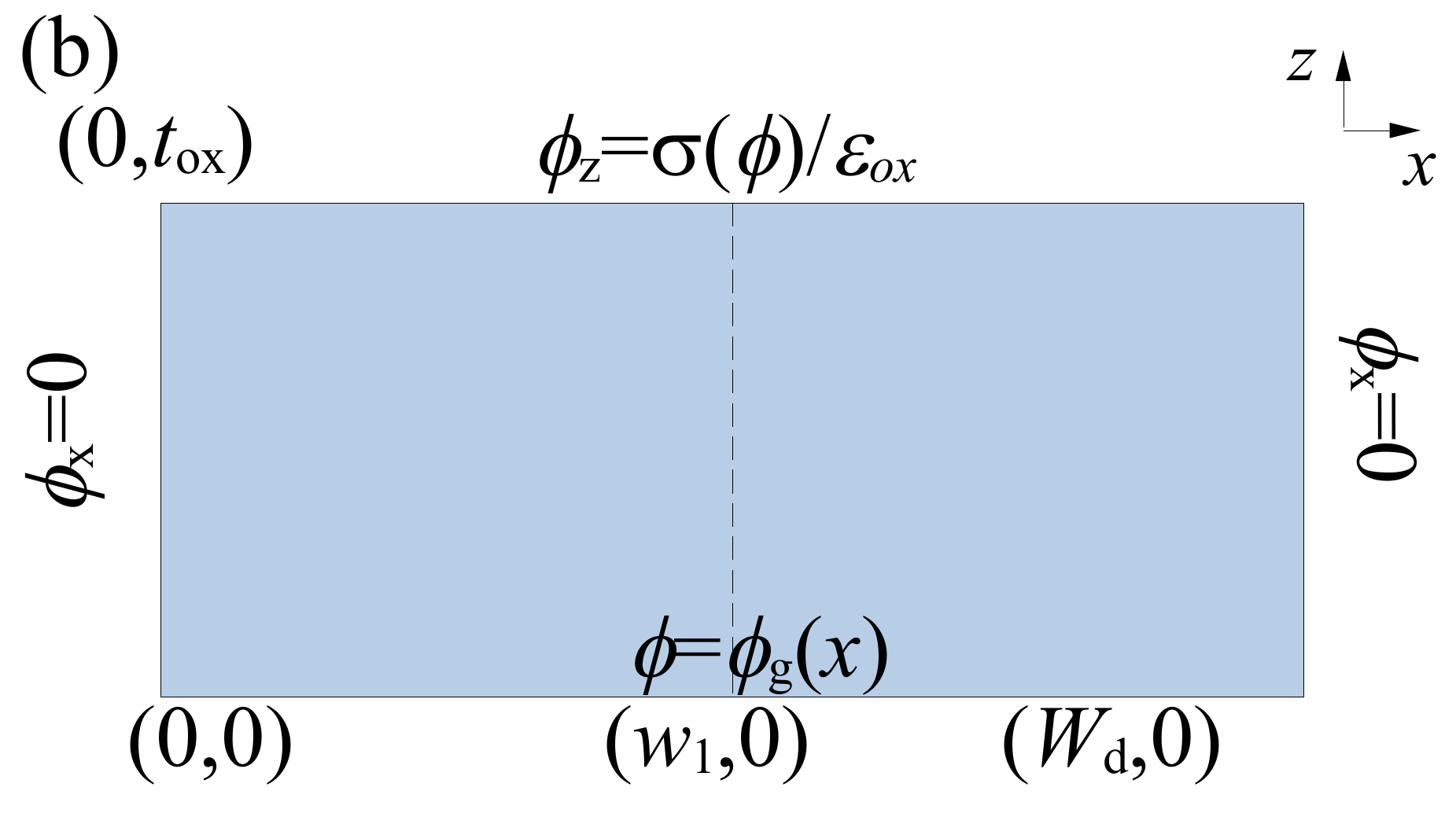}
 \caption{(a) Bottom: schematic cross section of a gate tunable 2D \textit{pn}-junction (not drawn to scale). Top: sketch of the device band diagram showing the depletion width. (b) Sketch of a symmetric depletion region showing critical coordinates and the boundary conditions required for solving the Poisson's equation within it, where $\phi_{\rm x,z} = \partial \phi / \partial x,z$. For an asymmetric depletion region, $w_1$ is not the middle point in the \textit{x}-direction.}
\label{fig:device}
\end{figure}

The electrostatics of a GT 2D junction depicted in Fig. \ref{fig:device}(a) can be made analytically treatable under the depletion approximation. Specifically, the electrostatic potential in the depletion region of the 2D semiconductor without chemical doping, as the one considered here, comes from the solution of the 2D Poisson equation under the assumption that there is no mobile charge inside. Fig. \ref{fig:device}(b) shows a sketch of such a region with the boundary conditions assumed in this work. Outside of the depletion region, the electrostatic potential at each side of the \textit{pn} junction, far away from the depletion region, is considered a constant and equal to the lowest and highest of the potential inside of the depletion region, those extreme values reached at the \textit{p}- and \textit{n}-boundary of the quasi neutral region, respectively. For this study, the physical gap between bottom-gates, $l_{\rm gap}$, has been considered to be much shorter than the depletion width $W_{\rm d}$ and hence, it has been neglected. The latter consideration has no impact on the results here as discussed below (cf. section III) and shown elsewhere \cite{ChaJim21}. The device width along the \textit{y}-axis is large enough to consider that the junction is uniform in that direction. Thus, models presented here do not depend on the device width. Furthermore, an infinitesimally thin semiconductor considered here reduces the 2D Poisson equation to the Laplace's equation within the total computing region constrained by the following general boundary conditions: homogeneous Neumann boundary conditions at $x=0$ and $x=W_{\rm d}$ forcing a zero electric field outside the depletion region; Dirichlet boundary conditions $\phi = \phi_{\rm g}(x)$ at the bottom-gate contacts ($z=0$) whereas a non-homogeneous Neumman condition $\phi_{\rm z}= \sigma(\phi)/\epsilon_{\rm ox}$ has been set at the semiconductor plane ($z=t_{\rm ox}$), where $\sigma=q(p-n)$ is the charge density of the 2D semiconductor with $p(n)$ as the hole( (electron) carrier density, and $\epsilon_{\rm ox}$ the dielectric permittivity. Notice that $\sigma$ within the depletion region is zero for the 2D semiconductor without chemical doping as the one considered here.

The electrostatic potential profile $\phi(x,z)$, obtained from the solution of the 2D Poisson's equation within the computing region (cf. Fig.\ref{fig:device}(b)), can be expressed as (see Appendix A for details on the solving process)

\begin{equation}
\phi(x,z) = \sum_{\rm k=1}^\infty \left\lbrace A_{\rm k} \cos \left(\lambda_{\rm k} x\right) \cosh\left[\lambda_{\rm k} (t_{\rm ox}-z)\right]\right\rbrace,
\label{eq:potential_gral}
\end{equation}

\noindent where

\begin{equation}
A_{\rm k} = \frac{2}{\lambda_{\rm k}W_{\rm d}} \frac{\sin\left(\lambda_{\rm k}w_1\right)\left(\phi_1 - \phi_2\right)}{\cosh\left(\lambda_{\rm k}t_{\rm ox}\right)},
\label{eq:coeff}
\end{equation}

\noindent with $\rm k = 1, 2, 3, \ldots$, $\lambda_{\rm k} = \left(\rm k\pi\right)/W_{\rm d}$, $t_{\rm ox}$ the oxide thickness, $w_1$ the splitting point between the two gates where the electrostatic potential changes from $\phi_{1}= \mathit{V}_{\rm g1}$ in the gate 1 to $\phi_{2}= \mathit{V}_{\rm g2}$ in the gate 2. It can be inferred that for a symmetric electrostatic doping ($V_{\rm g1}=-V_{\rm g2}$), $w_1 = W_{\rm d}/2$.

Eqs. (\ref{eq:potential_gral}) and (\ref{eq:coeff}) provide a general solution of the electrostatic potential of GT 2D junctions for both symmetrical ($\phi_1=-\phi_2$) and asymmetrical ($\phi_1\neq -\phi_2$) electrostatic doping, provided that the depletion region parameters, namely, $W_{\rm d}$ and $w_{1}$, could be previously obtained. The latter parameters are obtained by considering the following: \textit{(i)} the electrostatic potential is known at a quasi-neutral region along the 2D semiconductor, $\phi_{\rm o1}=\phi(0,t_{\rm ox})$, calculated at either the \textit{p}- or \textit{n}-type region (e.g., \textit{p}-type), and \textit{(ii)} that $\phi_1(0,0)=V_{\rm g1}$.

For \textit{(i)}, an analysis of the 1D electrostatics in the \textit{z}-direction is required. $\phi_{\rm o1(o2)}$ can be estimated from an analytical solution of a 1D metal-oxide-semiconductor (MOS) model obtained by considering the band profile shown in Fig. \ref{fig:1Dmos} and analyzed in detail in Appendix B. 

\begin{figure}[htb!]
\begin{center}
 \includegraphics[width=0.5\textwidth]{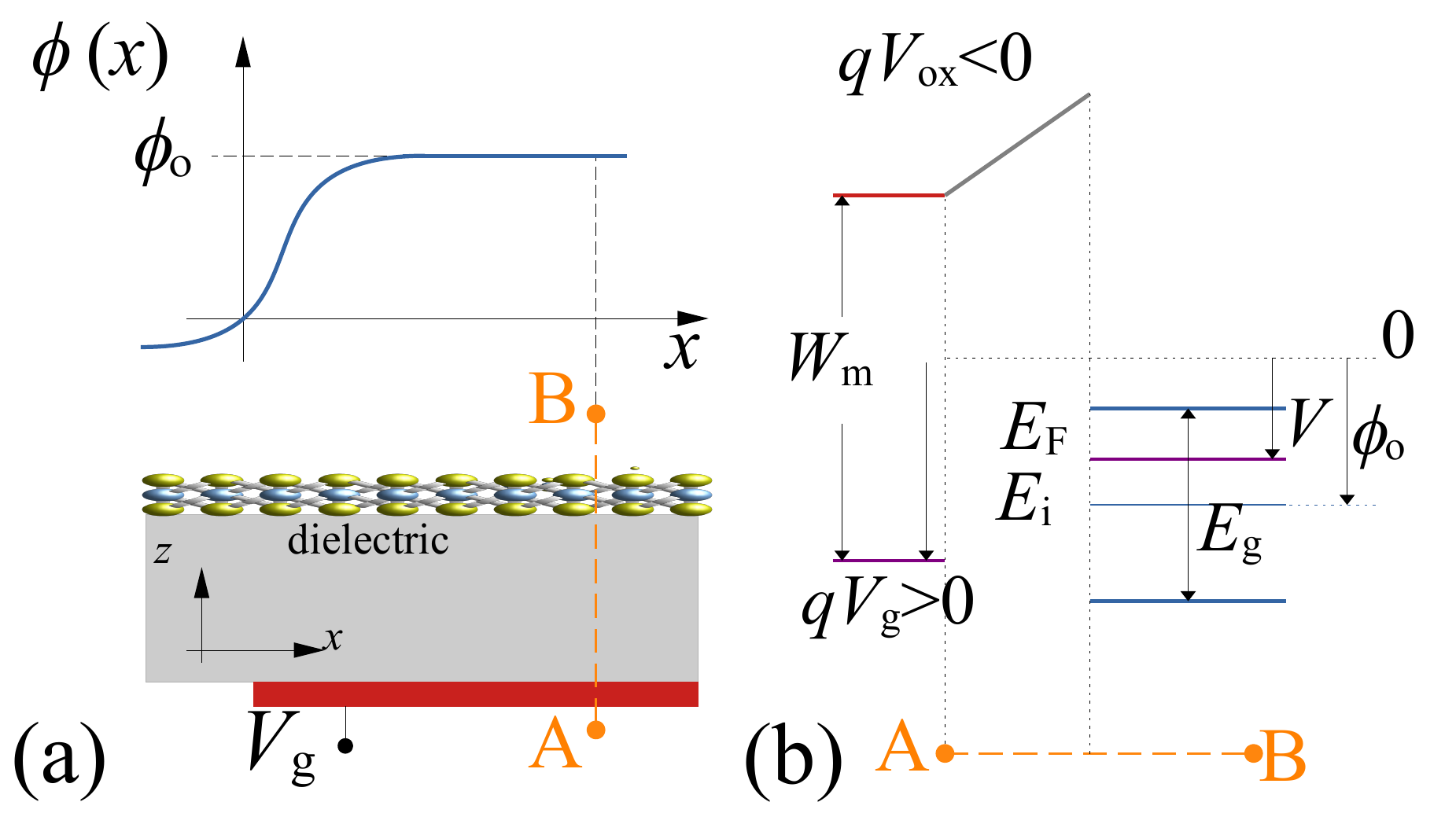}
 \caption{(a) Bottom: a cross-section of the GT 2D \textit{pn}-junction showing only one gate. Top: sketch of the electrostatic potential profile over the \textit{x}-direction along the semiconductor. (b) Band profile of the metal-oxide-semiconductor structure considering the layers across the cut AB shown in the bottom part of (a).}
\label{fig:1Dmos}
\end{center}
\end{figure}

\begin{figure}[!htb]
\centering
\includegraphics[height=0.21\textwidth]{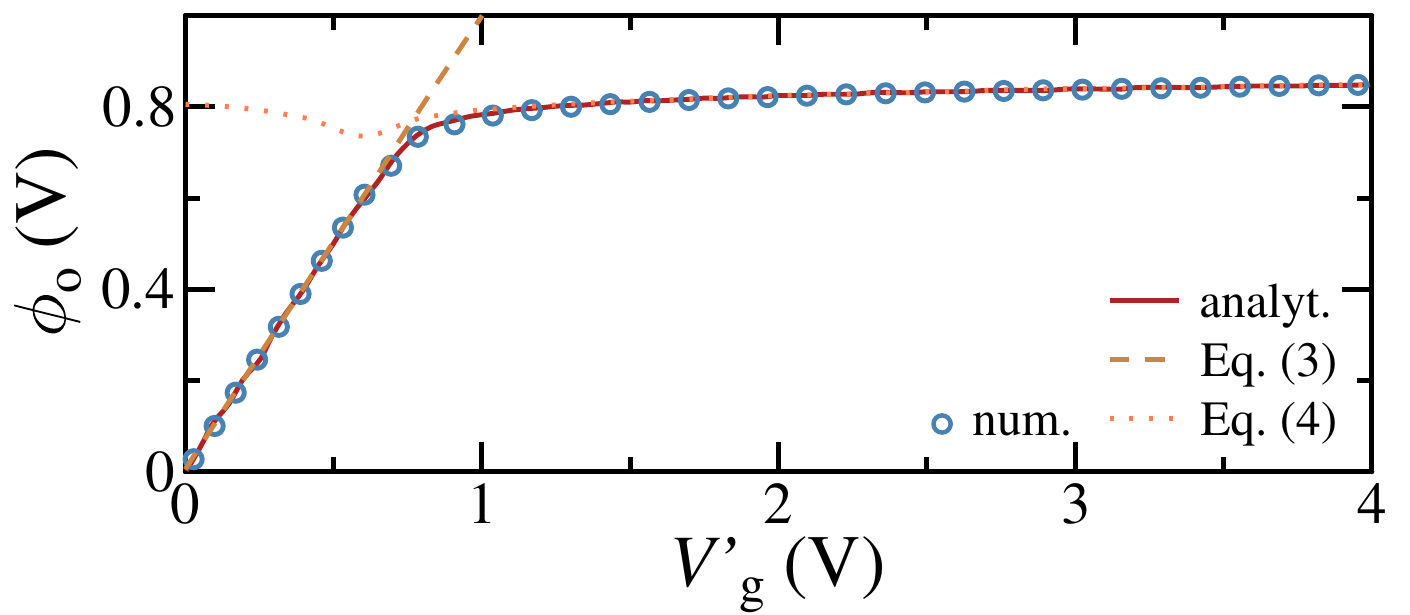} \vspace{-0.5cm}
\caption{$\phi_{\rm o}$ versus $V_{\rm g}'$: numerical simulations (symbols) and analytical results with Eq. (\ref{eq:phio_model}) (solid line). The model parameters are $E_{\rm g}=\SI{1.8}{\electronvolt}$, $t_{\rm ox}=\SI{300}{\nano\meter}$, $\epsilon_{\rm ox}=\SI{3.9}{}\epsilon_0$ and $\mu=\SI{1.003}{}$.}
\label{fig:phio_valid}
\end{figure}

The 1D MOS model \cite{ChaJim21} results in a non-linear equation for $\phi_{\rm o}$ (cf. Eq. (\ref{eq:appB2})), whose solution can be expressed by analytical piecewise functions. By considering an overdrive gate voltage $V_{\rm g}'$ embracing flat-band conditions (see definition in Appendix B), for a $V_{\rm g}' \lesssim V_{\rm th}$,

\begin{equation}
\phi_{\rm o<} = V_{\rm g}',
\label{eq:phio_izq}
\end{equation}

\noindent and

\begin{equation}
\phi_{\rm o>} = \frac{E_{\rm g}}{2q} + \frac{kT}{q} \log \left\lbrace\exp \left[\frac{(V_{\rm g}' - V_{\rm th})C_{\rm ox}}{q n_{\rm 0}} \right]-1 \right\rbrace,
\label{eq:phio_der}
\end{equation}

\noindent for $V_{\rm g}'>V_{\rm th}$, with the 2D semiconductor bandgap $E_{\rm g}$, the Boltzmann constant $k$, the absolute temperature $T$, the electric charge $q$, the oxide capacitance $C_{\rm ox}$, $n_{\rm 0}=g_{\rm 2D}kT$ with the band-edge density of states $g_{\rm 2D}$ (see definition in Appendix B) and the threshold voltage $V_{\rm th}$ defined as the \SI{60}{}\% of $E_{\rm g}/(2q)$  \cite{ChaJim21}. In order to obtain a smooth and continous analytical function for any gate voltage, an educated solution combining Eqs. (\ref{eq:phio_izq}) and (\ref{eq:phio_der}) yields 

\begin{equation}
\phi_{\rm o} = \phi_{\rm m} - \sqrt{\phi_{\rm m}^2 - \phi_{\rm o>}\phi_{\rm o<}},
\label{eq:phio_model}
\end{equation}

\noindent where 

\begin{equation}
\phi_{\rm m} = \frac{1}{2} \left(\phi_{\rm o>} + \mu \phi_{\rm o<}\right),
\end{equation}

\noindent with $\mu$ as a fitting parameter. As shown in Fig. \ref{fig:phio_valid}, the proposed analytical solution matches with the numerical device simulations of the 1D MOS model considering MoS$_2$ as the 2D semiconductor.

By applying the conditions \textit{(i)} and \textit{(ii)} into Eq. (\ref{eq:potential_gral}), it is found that

\begin{equation}
\phi_{\rm o1} = \sum_{\rm k=1}^\infty \left[\frac{2}{\pi \rm k} \frac{\sin\left(\pi{\rm k}\frac{w_1}{W_{\rm d}}\right)\left(\phi_1 - \phi_2\right)}{\cosh\left(\pi{\rm k}\frac{t_{\rm ox}}{W_{\rm d}}\right)}\right],
\label{eq:phi_o1}
\end{equation}

\noindent and

\begin{equation}
\phi_{\rm 1} = \sum_{\rm k=1}^\infty \left[\frac{2}{\pi\rm k} \sin\left(\pi{\rm k}\frac{w_1}{W_{\rm d}}\right)\left(\phi_1 - \phi_2\right) \right],
\label{eq:phi_1}
\end{equation}

\noindent which are non-linear equations for the variables $W_{\rm d}$ and $w_1$. 

From Eq. (\ref{eq:phi_1}) and by considering the convergence of the series $\sum_{\rm k=1}^\infty\left[(1/(\pi \rm k))\sin (\pi \rm k \mathit{x})\right] \rightarrow (-1/2)(x-1)$ $\forall$ $x<1$, the following ratio is obtained

\begin{equation}
\frac{w_1}{W_{\rm d}} = 1 - \frac{\phi_1}{\phi_1-\phi_2},
\label{eq:ratio1}
\end{equation}

\noindent from which it can be seen that for the symmetric case, i.e., $\phi_1=-\phi_2$, $w_1=W_{\rm d}/2$ holds. By replacing Eq. (\ref{eq:ratio1}) in Eq. (\ref{eq:phi_o1}), the following equation results

\begin{equation}
-\frac{1}{2}\frac{\phi_{\rm o1}}{\phi_2} = \sum_{\rm k=1}^\infty \left[\frac{1}{\pi {\rm k} r } \frac{\sin\left(\pi{\rm k}r \right)}{\cosh \left(\pi {\rm k} \frac{t_{\rm ox}}{W_{\rm d}} \right)}\right],
\label{eq:eq7}
\end{equation}

\noindent with $r=\phi_2/(\phi_2 - \phi_1)$. Eq. (\ref{eq:eq7}) can be solved numerically to obtain $W_{\rm d}$ and, consequently, $w_1$ from Eq. (\ref{eq:ratio1}).

Alternatively, a general analytical solution to obtain $W_{\rm d}$ is proposed in this work by considering the first term ($\rm k = 1$) of the sum in Eq. (\ref{eq:eq7}), which leads to

\begin{equation}
%W_{\rm d} = \pi t_{\rm ox} \left\lbrace \ln \left[ -\frac{\phi_2}{\phi_{\rm o1}} \rm {sinc}\mathit{ \left( \frac{r}{\pi} \right)} \right] \right\rbrace^{-1},
W_{\rm d} = \frac{\pi t_{\rm ox}}{\rm{sech}^{-1} \left[ -\frac{1}{2\rm {sinc} (\mathit{r})}\frac{\phi_{\rm o1}}{\phi_{\rm 2}} \right]},
\label{eq:Wd} 
\end{equation}

\noindent being sinc($r$) the normalized sinc function. Eq. (\ref{eq:Wd}) is valid for all values of $r$ between \SIlist{0;1}{} as long as $t_{\rm ox}/W_{\rm d} \geq 0.3$ (cf. Appendix C).

For the case of a GT 2D junction with symmetrical applied gate voltages, the depletion width is symmetrical. Hence, after some algebra, an expression for the depletion width for this case ($r=1/2$) obtained from Eq. (\ref{eq:eq7}) reads

\begin{equation}
W_{\rm d}= \frac{\pi t_{\rm ox}}{{\rm {sech}} ^{-1}\left(-\frac{\pi}{4} \frac{\phi_{\rm o1}}{\phi_2} \right)}\Bigg\rvert_{\phi_1=-\phi_2},
\label{eq:Wd_1}
\end{equation}

\noindent where in contrast to a model for chemically doped 2D \textit{pn}-junctions suggested elsewhere  \cite{AchYes02,IlaAme18,ChaJim18} there is no dependence on the oxide dielectric constant but only on its thickness. A phenomenological expression previously presented in  \cite{ChaJim21} for $W_{\rm d}$ of 2D junctions in this same scenario, is a particular case of the general analytical solution presented here (cf. Eq. (\ref{eq:Wd_1})). The physics-based and straightforward $W_{\rm d}$ expressions obtained here can be used also to calculate transport-related parameters of 2D junctions  \cite{LowHon09,GhaKho10}, however, this is out of the scope of the present study. 

\section{Results and Discussion}

\begin{figure}[!htb]
\centering
\includegraphics[height=0.225\textwidth]{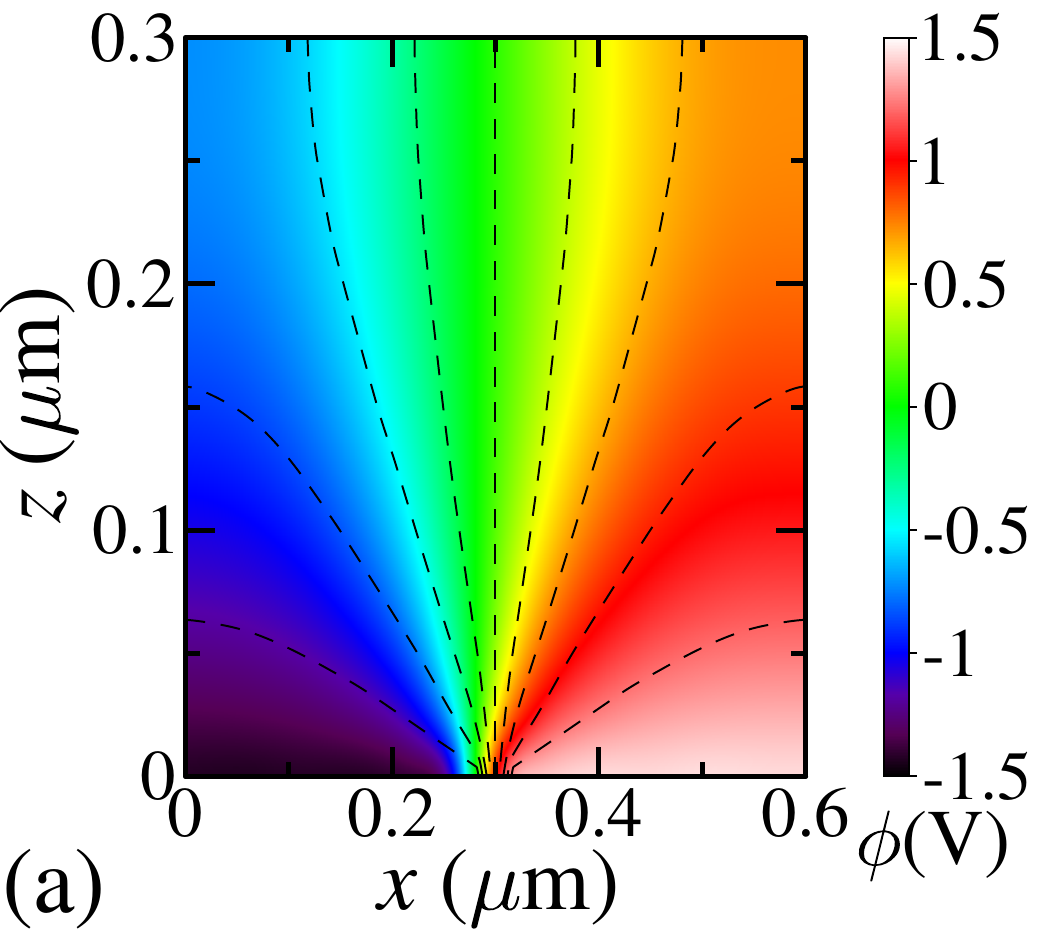}
\includegraphics[height=0.225\textwidth]{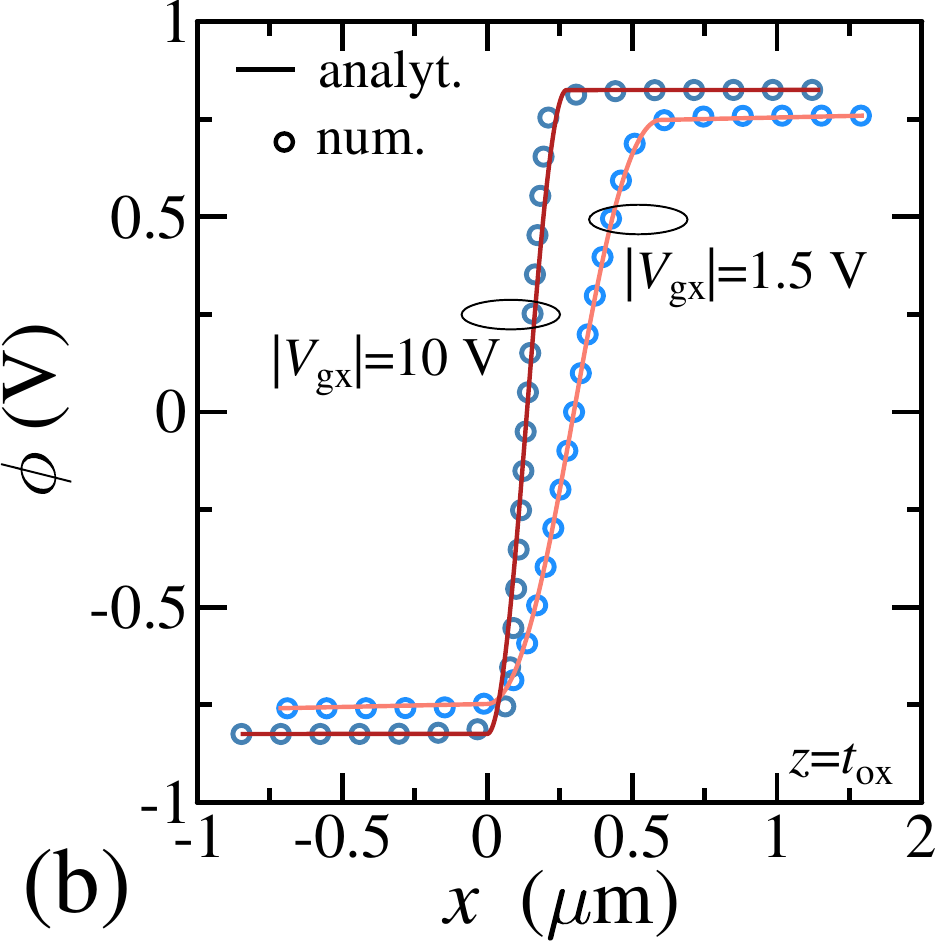}
\caption{Electrostatic potential of a 2D MoS$_2$ junction with symmetrical applied voltages. (a) Analytical results within the depletion region inside the oxide obtained with Eq. (\ref{eq:potential_gral}) for $V_{\rm g1}=-V_{\rm g2}=-\SI{1.5}{\volt}$. Dashed lines are the equipotential lines. (b) Analytical (lines) and numerical simulation results (symbols) along the \SI{2}{\micro\meter} 2D semiconductor for $V_{\rm g1}=-V_{\rm g2}=-\SI{1.5}{\volt}$ and -\SI{10}{\volt}.}
\label{fig:results_01}
\end{figure}

\begin{figure}[!htb]
\centering
\includegraphics[height=0.225\textwidth]{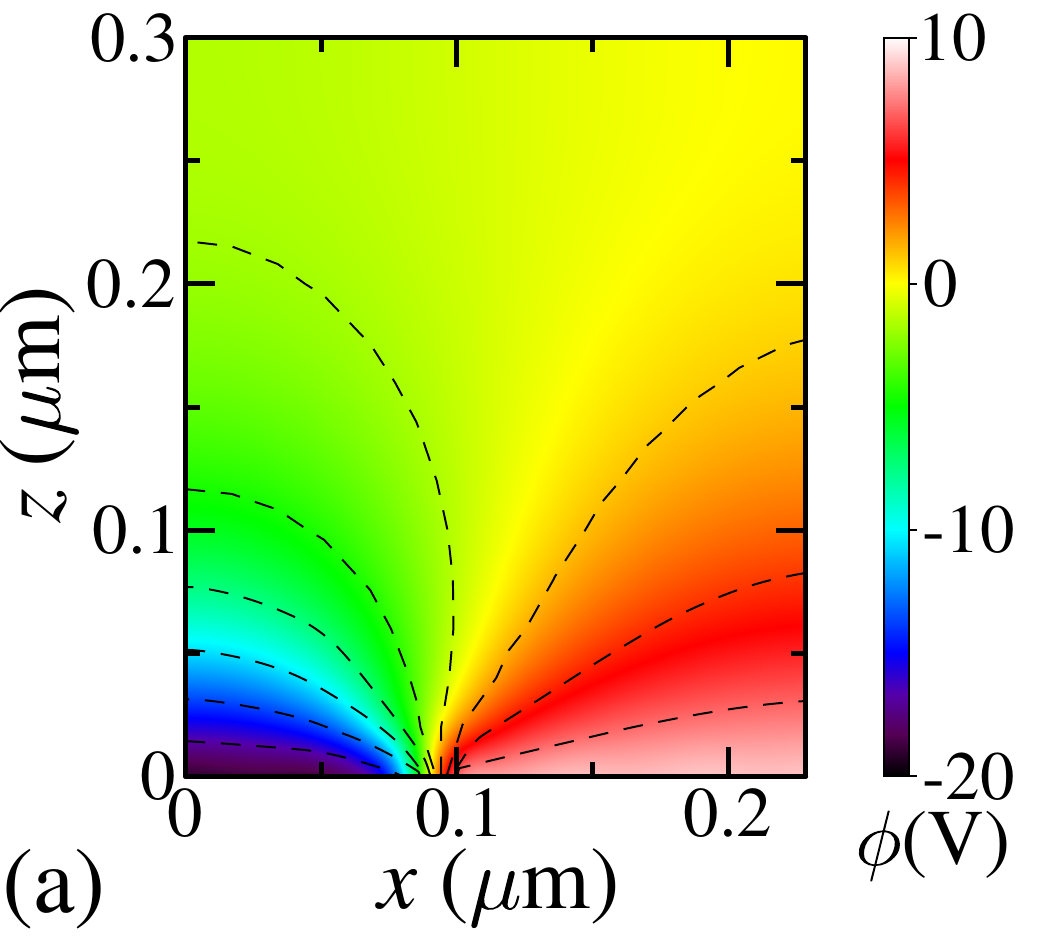}
\includegraphics[height=0.225\textwidth]{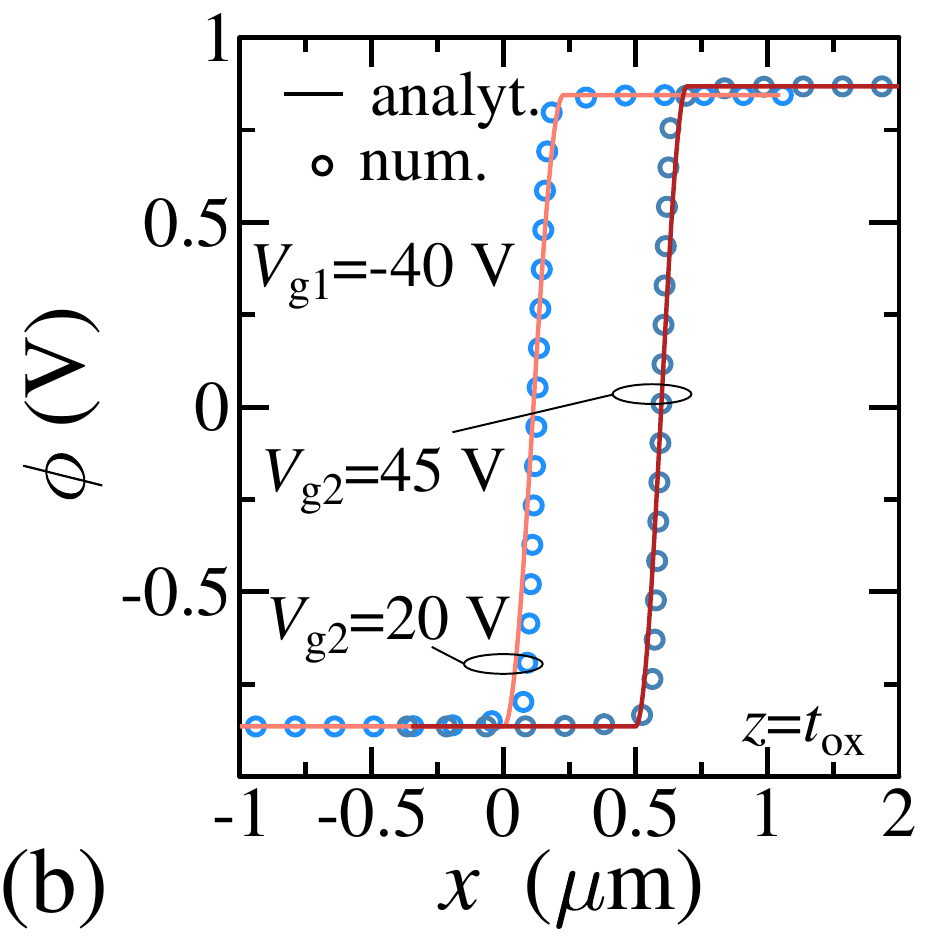} \\
\caption{Electrostatic potential of a 2D MoS$_2$ junction with asymmetrical applied voltages ($V_{\rm g1}\neq-V_{\rm g2}$). (a) Analytical results within the computational region obtained with Eq. (\ref{eq:potential_gral}) at $V_{\rm g1}=-\SI{20}{\volt}$ and $V_{\rm g2}=\SI{10}{\volt}$. (b) Analytical (lines) and numerical simulation results (symbols) along the 2D semiconductor at different gate voltages. Electrostatic potentials for the case with smallest asymmetry have been shifted to the right by an amount of \SI{0.5}{\micro\meter} for visualization purposes.}
\label{fig:results_03}
\end{figure}

\begin{figure}[!htb]
\centering
\includegraphics[height=0.225\textwidth]{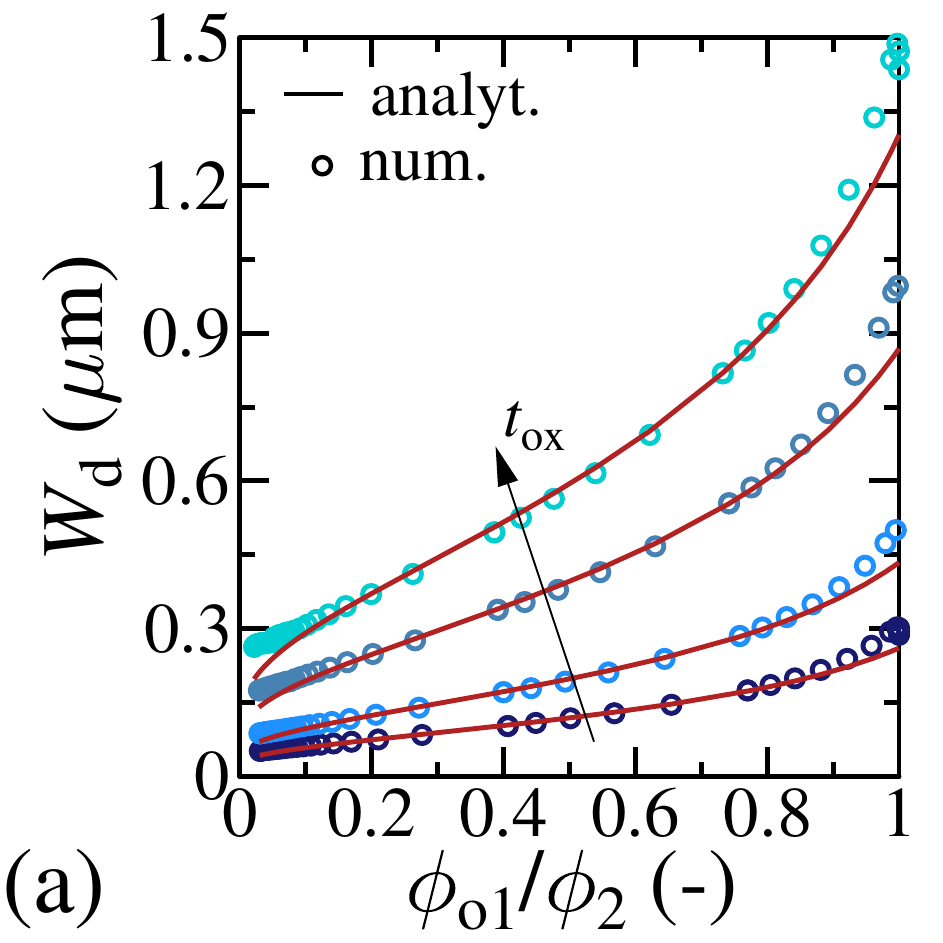}
\includegraphics[height=0.225\textwidth]{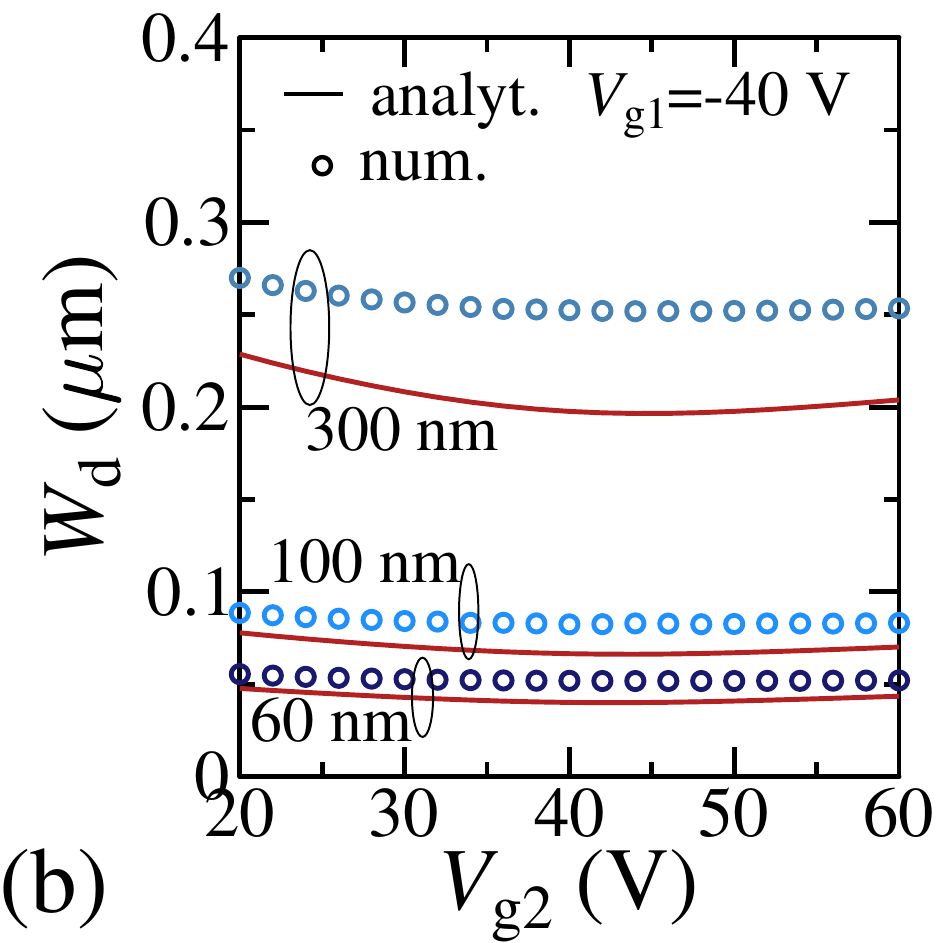} \\
\caption{Variation of the depletion width of (a) symmetric GT 2D junction with the ratio $\phi_{\rm o1}/\phi_2$ for $t_{\rm ox}$ equal to \SIlist{60;100;200;300}{\nano\meter}, and of (b) asymmetric GT 2D junction with $V_{\rm g2}$ for 3 different $t_{\rm ox}$. Symbols are numerical device simulation results and lines are results from the analytical model proposed here.}
\label{fig:results_02}
\end{figure}

The model presented here has been evaluated considering a \SI{2}{\micro\meter}-long MoS$_2$ lateral \textit{pn}-junction. Unless stated otherwise, a \SI{300}{\nano\meter}-thick SiO$_2$ oxide separates the back-gates and the 2D semiconductor. $l_{\rm gap}$ is considered \SI{0}{} as in the analytical model. Furthermore, previous studies have shown that $l_{\rm gap}$ defines mostly $W_{\rm d}$ by a linear relation for long enough values  \cite{ChaJim21}. The device has been studied in two different scenarios: with symmetric ($V_{\rm g1}=-V_{\rm g2}$) and asymmetric ($V_{\rm g1}\neq-V_{\rm g2}$) electrostatic doping. Hereinafter, $V_{\rm gx}$ indicates the overdrive gate voltage unless stated otherwise. Numerical device simulations (NDS) have been performed to benchmark the analytical approach presented here (cfs. Figs. \ref{fig:results_01}-\ref{fig:results_03}). $W_{\rm d}$ in the NDS tool corresponds to the semiconductor $V_{\rm g}$-induced charge $Q_{\rm sc}(x)$ is equal to 43.6\% of the induced charge at each side of induced charge regions of the junction. Further details on the experimentally-calibrated in-house physics-based simulation tool can be found elsewhere \cite{ChaJim21}. In Fig. \ref{fig:results_01} and Fig. \ref{fig:results_03} both the 2D electrostatic potential inside the dielectric and the 1D electrostatic potential in the semiconductor for the symmetric and asymmetric cases are shown, respectively. Fig. \ref{fig:results_02} shows the depletion width dependence on the gate voltages for some symmetric and asymmetric cases with different values of $t_{\rm ox}$.  

For the symmetric case, gate voltages of $\pm$\SI{1.5}{\volt} and $\pm$\SI{10}{\volt} have been used, yielding $r=1/2$ for each electrostatic doping. The symmetric electrostatic potential (cf. Eq. (\ref{eq:potential_gral})) has been obtained with the model within the device depletion region for both inside the oxide and at the 2D semiconductor, provided $W_{\rm d} (\approx\SI{0.6}{\micro\meter})$ has been calculated with Eq. (\ref{eq:Wd_1}), as shown in Figs. \ref{fig:results_01}(a) for $V_{\rm g1}=-V_{\rm g2}=-\SI{1.5}{\volt}$, and (b) for both gate voltages, respectively.  

The equipotential line at $x=w_1\approx\SI{0.3}{\micro\meter}$ in Fig. \ref{fig:results_01}(a) corresponds to $\phi=\SI{0}{}$. The symmetric distribution of the total electrostatic potential within the oxide can be observed along the \textit{x}-direction, i.e., $\phi(0\leq x \leq w_1,z)=-\phi(w_1\leq x \leq W_{\rm d},z)$. Interestingly, the electric field lines\footnote{Electric field lines are perpendicular to the equipotential lines.} in gate tunable lateral 2D \textit{pn}-junctions here studied are different to those generated in chemically doped lateral 2D \textit{pn}-junctions, as studied in refs. \cite{NipJay17}-\cite{ChaJim18} where there is a non-zero component of the electric field perpendicular to the semiconductor plane inside the dielectric next to the semiconductor in the depletion region. However, for GT devices, such electric field component is vanished, and hence, there is a very weak dependence of $W_{\rm d}$ (Eq. (\ref{eq:Wd})) on $\epsilon_{\rm ox}$ in contrast with the chemical doped case.

The analytical model successfully describes the symmetric electrostatic potential profile along the 2D semiconductor (at $z=t_{\rm ox}$) as shown in Fig. \ref{fig:results_01}(b) by comparing it with numerical simulation results of the same device at different bias. These curves exhibit depletion widths of $\sim \SI{0.6}{\micro\meter}$ and $\SI{0.27}{\micro\meter}$ for $V_{\rm g1}=-V_{\rm g2}=-\SI{1.5}{\volt}$ and $V_{\rm g1}=-V_{\rm g2}=-\SI{10}{\volt}$, respectively, which correspond to the calculated values by means of Eq. (\ref{eq:Wd_1}) with $\phi_{\rm o1} = -\SI{0.75}{\volt}$ and $-\SI{0.82}{\volt}$ from Eq. (\ref{eq:phio_model}).  The left ($x<0$ in the computational region) and right ($x>W_{\rm D}$) 2D semiconductor quasi-neutral regions are described by the minimum and maximum values of $\phi$ (obtained with the analytical approach used here, cf. Eq. (\ref{eq:phio_model})), respectively, in the corresponding depletion region.

For the asymmetric case, the same MoS$_2$-based \textit{pn} junction studied above ($t_{\rm ox}=\SI{300}{\nano\meter}$) has been biased at three differenct configurations. Fig. \ref{fig:results_03}(a) shows the potential contour plot with equipotential lines from the analytical model for the case with $V_{\rm g1}=-\SI{20}{\volt}$ and $V_{\rm g2}=\SI{10}{\volt}$ ($r=\SI{0.33}{}$) exhibiting $W_{\rm d}\approx\SI{0.23}{\micro\meter}$ and $w_1\approx\SI{0.09}{\micro\meter}$, as given by Eq. (\ref{eq:ratio1}). The different electrostatic doping induced by each gate breaks the symmetry of $\phi$ within the computational region as observed by the analytical modeling (cf. Eq.(\ref{eq:potential_gral})). A zero perpendicular electric field component next to the semiconductor can be observed, similar to the symmetric case, explaining the weak dependence of $W_{\rm d}$ on $\epsilon_{\rm ox}$ even in the asymmetric case. 

Fig. \ref{fig:results_03}(b) shows analytical and numerical results of the electrostatic potential along the 2D semiconductor of GT MoS$_2$ junction at two different asymmetric bias sets: $V_{\rm g1} = -\SI{40}{\volt}$; $V_{\rm g2} = \SI{20}{\volt}$ (strong asymmetry $r=0.33$) and for $V_{\rm g1} = -\SI{40}{\volt}$; $V_{\rm g2} = \SI{45}{\volt}$ (small asymmetry $r=0.53$) exhibiting predicted $W_{\rm d}$ equal to $\sim\SI{0.23}{\micro\meter}$ and $\sim\SI{0.20}{\micro\meter}$, respectively. The more asymmetric the electrostatic doping is the more distant is $w_1$ from $W_{\rm d}/2$. Hence, $\phi(x-w_1=0,t_{\rm ox})\neq\SI{0}{\volt}$ in contrast to the symmetric scenario.

The analytical depletion width model (cf. Eqs. (\ref{eq:Wd}) and (\ref{eq:Wd_1})) is able to describe NDS results of both symmetric and asymmetric GT MoS$_2$ junctions with different $t_{\rm ox}$. Fig. \ref{fig:results_02}(a) shows the analytical (cf. Eq. (\ref{eq:Wd_1})) and NDS results of $W_{\rm d}$ for symmetric \textit{pn}-junctions as a function of $\phi_{\rm o1}/\phi_2$. As example cases, the ratio takes values of \SI{0.5}{} and \SI{0.082}{} for the two symmetrical devices described in Fig. \ref{fig:results_01}(b).

The depletion widths obtained with the analytical model (cf. Eq. (\ref{eq:Wd})) for asymmetric GT MoS$_2$ junctions with different $t_{\rm ox}$ are compared to NDS results in Fig. \ref{fig:results_02}(b). The electrostatic doping is enabled by applying $V_{\rm g1}=-\SI{40}{\volt}$ and sweeping $V_{\rm g2}$ from \SIrange{20}{60}{\volt}, yielding a minimum and maximum values of $r$ equal to \SI{0.3}{} and \SI{0.6}{}, respectively. The $V_{\rm g2}$-dependence of $W_{\rm d}$ at different $t_{\rm ox}$ is qualitatively captured by the model. The maximum relative error for the analytical results in comparison to the NDSs is of $\sim\SI{21}{}\%$ for the devices under study. The origin of this deviation is discussed next.  

$W_{\rm d}$ values from NDS could, in fact, be lower than the ones reported here in the asymmetric case due to the not unique definition of the computational space related to depletion region, specially its beginning and ending, i.e., the \textit{x}-limits. In this work, we have assumed the same criterium of the 43.6\% of the induced charge density in the quasi-neutral regions like in the symmetric case \cite{ChaJim18}, \cite{ChaJim21}. However, due to the large transition region separating the fully depletion region and the quasi-neutral depletions regions exhibited by 2D lateral \textit{pn}-junctions in comparison to 3D \textit{pn}-junctions, as discussed elsewhere \cite{ChaJim18}, this criterium could be relaxed for devices with asymmetric electrostatic doping and pristine 2D semiconductors towards an improved description of the NDS by the analytical model. Additionally, the 1D analytical model of $\phi_{\rm o1(o2)}$ might be another source of error since it only considers the impact of one gate voltage in contrast to the 2D numerical solution depending on both applied bias. By considering these discussions and by showing that all conditions are fulfilled, i.e., $0<r<1$ and $t_{\rm ox}/W_{\rm d}>\SI{0.3}{}$, the analytical model of $W_{\rm d}$ in the asymmetric scenario, can be used to set reliable minimum limit of values. 

\section{Conclusion}

The 2D Poisson equation has been solved analytically yielding models for the electrostatic potential and depletion width of GT 2D lateral \textit{pn}-junctions with symmetric and asymmetric electrostatic doping applied by two separated bottom-gates. The analyitical $\phi$ model enables to elucidate the electrostatic potential within both the oxide and the 2D semiconductor, regardless the symmetric bias conditions. The analytical $\phi$ and $W_{\rm d}$ models have been benchmarked against numerical device simulations of GT MoS$_2$-based junctions. For the symmetric electrostatic doping scenario ($V_{\rm g1}=-V_{\rm g2}$), the physics-based analytical model is able to describe the NDS results within the 2D semiconductor at all bias and for different oxide thickness; whereas for the asymmetric scenario ($V_{\rm g1}\neq-V_{\rm g2}$), the model captures the bias dependence of practical cases for different oxide thicknesses while yielding reliable $W_{\rm d}$ minimum values within a confidence range of $\gtrsim\SI{80}{}\%$. The straightforward analytical models proposed here for the electrostatics of GT 2D junctions intend to be of aid for technology improvements by unveiling some aspects of the internal physical mechanisms as well as for the modeling community by considering the expressions provided here in electrostatics-dependent transport models.

\section*{Appendix A. Solution of the 2D Poisson equation}

The electrostatic potential within the depletion region can be expressed as the product of an \textit{x}-dependent function $X(x)$ and a \textit{z}-dependent function $Z(z)$, i.e., $\phi(x,z)=XZ$. Hence, by separation of variables, the Laplace's equation ($\nabla^2 \phi=0$) reads

\setcounter{equation}{0}
\renewcommand{\theequation}{\Alph{equation}.1}

\begin{equation}
Z \frac{\partial^2 X}{\partial x^2} + X \frac{\partial^2 Z}{\partial z^2}=0,
\label{eq:app1}
\end{equation}

\noindent from which a differential partial equation system can be obtained by dividing Eq. (\ref{eq:app1}) by $XZ$ such as

\setcounter{equation}{0}
\renewcommand{\theequation}{\Alph{equation}.2}

\begin{subequations}
\begin{eqnarray}
\frac{\partial^2 X}{\partial x^2} + \lambda^2 X = 0, \\
\frac{\partial^2 Z}{\partial z^2} - \lambda^2 Z = 0,
\end{eqnarray}
\label{eq:app_2}
\end{subequations}

\noindent with

\setcounter{equation}{0}
\renewcommand{\theequation}{\Alph{equation}.3}

\begin{equation}
\lambda^2 = -\frac{1}{X}\frac{\partial^2 X}{\partial x^2} = \frac{1}{Z}\frac{\partial^2 Z}{\partial z^2}.
\end{equation}

The solutions of Eq. (\ref{eq:app_2}) can be expressed in a general form as

\setcounter{equation}{0}
\renewcommand{\theequation}{\Alph{equation}.4}

\begin{subequations}
\begin{eqnarray}
X(x) = a_{\rm k} \sin\left(\lambda_{\rm k}x\right) + b_{\rm k} \cos\left(\lambda_{\rm k}x\right), \\
Z(z) = c_{\rm k} \exp\left(-\lambda_{\rm k}z\right) + d_{\rm k} \exp\left(\lambda_{\rm k}z\right),
\end{eqnarray}
\label{eq:app_01}
\end{subequations}

\noindent for $\lambda^2>0$ and

\setcounter{equation}{0}
\renewcommand{\theequation}{\Alph{equation}.5}

\begin{subequations}
\begin{eqnarray}
X(x) = a_{\rm k} \exp\left(-\lambda_{\rm k}x\right) + b_{\rm k} \exp\left(\lambda_{\rm k}x\right), \\
Z(z) = c_{\rm k} \sin\left(\lambda_{\rm k}z\right) + d_{\rm k} \cos\left(\lambda_{\rm k}z\right),
\end{eqnarray}
\label{eq:app_011}
\end{subequations}

\noindent for $\lambda^2<0$. In this work, the former case ($\lambda^2>0$) has been considered without loss of generality.

By using Eq. (\ref{eq:app_01}a) and the Neumann boundary conditions $\phi_{\rm x}(x=0)=0$ and $\phi_{\rm x}(x=W_{\rm d})=0$ (cf. Fig. \ref{fig:device}(b)) lead to find $a_{\rm k}=0$ from the former boundary condition and consequently, $\sin(\lambda_{\rm k}W_{\rm d})=0$ for the second one, from which the $\lambda={\rm{k}}\pi/W_{\rm d}$ is obtained where $\rm{k}=1, 2, 3, \ldots$. Therefore,

\setcounter{equation}{0}
\renewcommand{\theequation}{\Alph{equation}.6}

\begin{equation}
X(x) = b_{\rm k}\cos\left(\frac{{\rm{k}}\pi}{W_{\rm d}}x\right).
\label{eq:x}
\end{equation}

Similarly, by considering Eq. (\ref{eq:app_01}b), the evaluation of $\phi_{\rm z}=0$ at $z=t_{\rm ox}$ yields $c_{\rm k}=d_{\rm k}\exp(2\lambda_{\rm k}t_{\rm ox}$), leading to

\setcounter{equation}{0}
\renewcommand{\theequation}{\Alph{equation}.7}

\begin{equation}
Z(z) = 2 d_{\rm k} \exp(\lambda_{\rm k}t_{\rm ox})\cosh \left[\lambda_{\rm k}(t_{\rm ox}-z)\right].
\label{eq:z}
\end{equation}

By using Eqs. (\ref{eq:x}) and (\ref{eq:z}) in the proposed definition of the electrostatic potential, Eq. (\ref{eq:potential_gral}) is found, whereas the definition of $A_{\rm k}$ is obtained as follows.

The value of the electrostatic potential at $z=0$ along the $x$-direction $\phi(x,0)=\phi_{\rm g}(x)$ reads as

\setcounter{equation}{0}
\renewcommand{\theequation}{\Alph{equation}.8}

\begin{equation}
\phi_{\rm g}(x) = \sum_{\rm k=1}^\infty \left[A_{\rm k} \cos(\lambda_{\rm k} x)  \cosh (\lambda_{\rm k} t_{\rm ox})\right].
\label{eq:phig1}
\end{equation}

\noindent Solving Eq. (\ref{eq:phig1}) for $A_{\rm k}$ leads to

\setcounter{equation}{0}
\renewcommand{\theequation}{\Alph{equation}.9}

\begin{equation}
A_{\rm k} = \frac{\int_0^{W_{\rm d}}\phi_{\rm g}(x)\cos(\lambda_{\rm k}x){\rm{d}}x}{\cosh(\lambda_{\rm k}t_{\rm ox})\int_0^{W_{\rm d}}\cos^2(\lambda_{\rm k} x){\rm{d}}x}.
\label{eq:Ak}
\end{equation}

\noindent The integral in the denominator yields $W_{\rm d}/2$. Since $\phi_{\rm g}(x) = \phi_1$ for $x<w_1$ and $\phi_{\rm g}(x) = \phi_2$ for $x>w_1$, the integral in the numerator can be split in a sum of integrals such as $\int_{0}^{W_{\rm d}}=\int_{0}^{w_{\rm 1}} + \int_{w_1}^{W_{\rm d}}$. Hence, 

\setcounter{equation}{0}
\renewcommand{\theequation}{\Alph{equation}.10}

\begin{subequations}
\begin{eqnarray}
\int_{0}^{w_1} \phi_{1} \cos(\lambda_{\rm k} x) {\rm{d}}x = \frac{\phi_1}{\lambda_{\rm k}}\sin(\lambda_{\rm k} w_1), \\
\int_{w_1}^{W_{\rm d}} \phi_{2} \cos(\lambda_{\rm k} x) {\rm{d}}x = -\frac{\phi_2}{\lambda_{\rm k}}\sin(\lambda_{\rm k} w_1).
\end{eqnarray}
\end{subequations}

Eq. (\ref{eq:coeff}) is hence determined by substituing the solutions of the integrals in Eq. (\ref{eq:Ak}).

\section*{Appendix B. 1D MOS model}

The band profile of a transversal 1D section of the GT 2D \textit{pn}-junction (Fig. \ref{fig:1Dmos}(a)) is shown in Fig. \ref{fig:1Dmos}(b). The section is chosen in a way that the Fermi level energy $E_{\rm F}$ considered is far away from the depletion region. From this band profile and by considering the charge conservation law, i.e., $Q_{\rm m}+Q_{\rm sc}=0$, and the voltage Kirchoff's law, i.e., $W_{\rm m}-qV_{\rm ox}=\chi+E_{\rm g}/2-q\phi_{\rm o}+qV_{\rm g}$, the following relation can be obtained

\setcounter{equation}{1}
\renewcommand{\theequation}{\Alph{equation}.1}

\begin{equation}
\left[\chi+E_{\rm g}/2 - q\phi_{\rm o} + qV_{\rm g} - W_{\rm m}\right]\frac{C_{\rm ox}}{q} + Q_{\rm sc}\left( \phi_{\rm o} - V \right) = 0
\label{eq:appB1}
\end{equation}

\noindent with the semiconductor electron affinity $\chi$, the band gap of the 2D semiconductor $E_{\rm g}$, the metal work function $W_{\rm m}$, the gate charge density $Q_{\rm m}$ and the $V_{\rm g}$-induced carrier charge density $Q_{\rm sc}(=\sigma)$ in the semiconductor which is a function of the local electrostatic potential $\phi_{\rm o}$ and of an electrochemical potential $V$ (arbitrarily referred).

In order to calculate $\phi_{\rm o}$, for instance in the \textit{n}-region and assuming thermal equilibrium with \newline$V=0$, $Q_{\rm sc}\approx -qn =-qn_{0}\log\left\lbrace 1+\exp\left[(q\phi_{\rm o}-E_{\rm g}/2)/(kT)\right]\right\rbrace$, where $n_0=g_{\rm 2D}kT=\left[g_{\rm v}g_{\rm s}m/(2\pi h^2)\right]kT$ defined by the band-edge effective mass, spin and valley degeneracy factors $m$, $g_{\rm v}$, $g_{\rm s}$ of the 2D semiconductor, respectively. Therefore, Eq. (\ref{eq:appB1}) can be written as

\setcounter{equation}{1}
\renewcommand{\theequation}{\Alph{equation}.2}

\begin{equation}
\begin{aligned}
\left( -\phi_{\rm o} + V_{\rm g}' \right)& C_{\rm ox} - \\& \hspace{-0.5cm}q n_{\rm 0} \log \left\lbrace 1 + \exp \left[\left(q \phi_{\rm o}-E_{\rm g}/2\right)/(kT)\right]\right\rbrace = 0,
\end{aligned}
\label{eq:appB2}
\end{equation}

\noindent which is a transcendental equation for $\phi_{\rm o}$ with $qV_{\rm g}'=qV_{\rm g}-W_{\rm m} + \chi +E_{\rm g}/2$, where the term $W_{\rm g}-\chi-E_{\rm g/2}$ describes the flat-band voltage. A piecewise analytical solution for Eq. (\ref{eq:appB2}) is given by Eqs. (\ref{eq:phio_izq})-(\ref{eq:phio_model}) in the main text. The model works regardless the device geometry and dielectric properties since in the hypothetical worst case where $C_{\rm ox}=-qn_{0}/(V_{\rm g}'-V_{\rm th})$, i.e., $\phi_{o>}=E_{\rm g}/(2q)$ (cf. Eq. (\ref{eq:phio_der})), a solution still exists for Eq. (\ref{eq:phio_model}).

For the analytical results presented in this work, the following parameter values have been considered: $T=\SI{300}{\kelvin}$, $g_{\rm v} = g_{\rm s} = 2$, $E_{\rm g} = \SI{1.8}{\electronvolt}$ and $m=\SI{0.57}{}m_0$ where $m_0$ is the free electron mass. The same values have been used in the numerical device simulations in \cite{ChaJim21}.

\section*{Appendix C: Validity of conditions for the analytical $W_{\rm d}$ model}

\begin{figure}[!htb]
\centering
\includegraphics[height=0.245\textwidth]{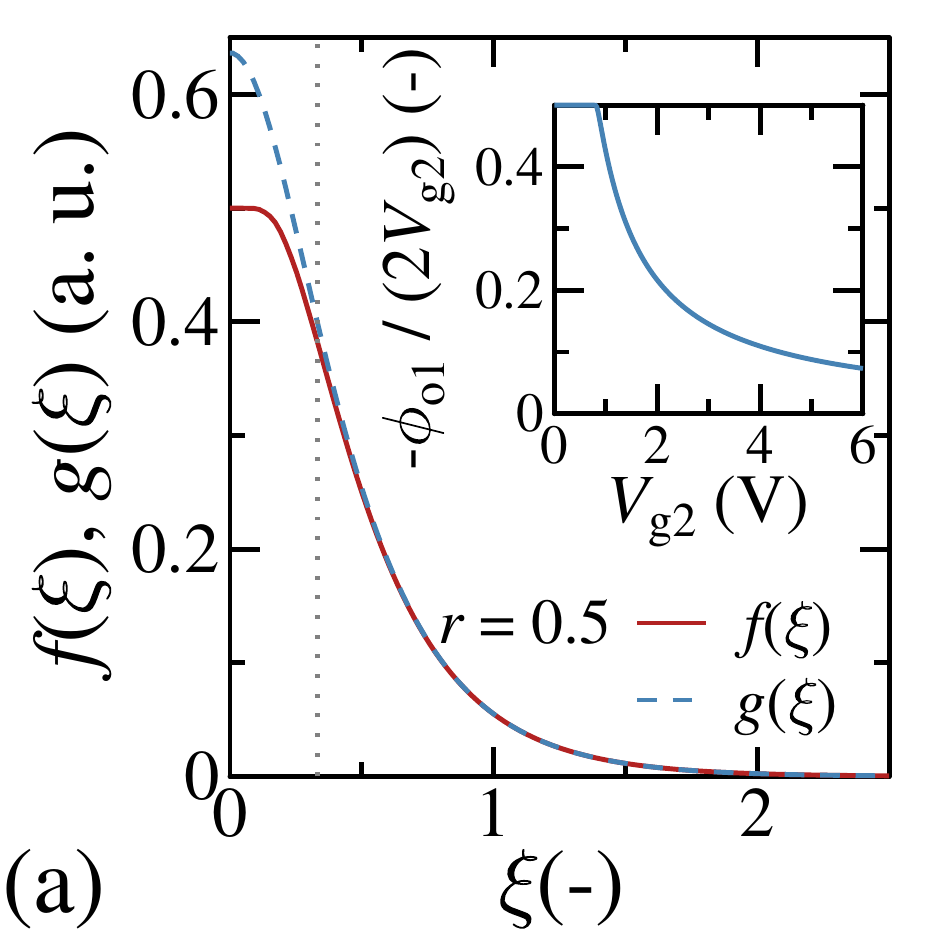} \hspace{-0.3cm}
\includegraphics[height=0.245\textwidth]{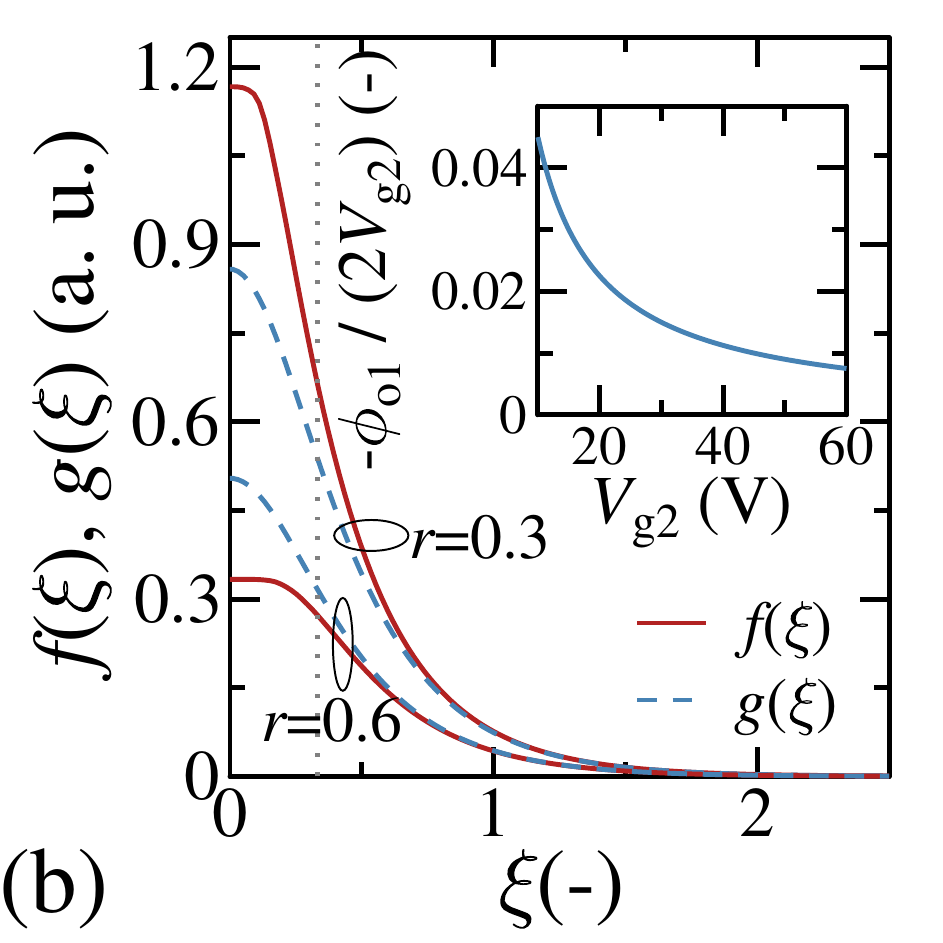}
\caption{Right side of Eq. (\ref{eq:eq7}) calculated with one (k=1) and several terms (k from 1 to 1200) terms for (a) symmetric and (b) asymmetric electrostatic doping. Insets: results of the left side of Eq. (\ref{eq:eq7}). Dotted line indicates $\xi=\SI{0.3}{}$ and is added as a guide for the eye. Values considered for this study are $t_{\rm ox}=\SI{300}{\nano\meter}$ for all cases whereas $\phi_{\rm o1}$ is equal to \SI{0.89}{} for the asymmetric case.}
\label{fig:appC1}
\end{figure}

Fig. \ref{fig:appC1} shows results of Eq. (\ref{eq:eq7}) as a function of $\xi=t_{\rm ox}/W_{\rm d}$ with $\rm k$-terms. For the symmetric case (Fig. \ref{fig:appC1}(a)), $\phi_{\rm o1}$ depends on $V_{\rm g1} = V_{\rm g2}$ whereas, for the asymmetric case (Fig. \ref{fig:appC1}(b)), $\phi_{\rm o1}$ is constant since $V_{\rm g1}$ is fixed and $V_{\rm g2}$ varies.

$f(\xi)$ and $g(\xi)$ correspond to the summatory on the right side of Eq. (\ref{eq:eq7}) calculated with one (k=1) and several terms (k from 1 to 1200), respectively. It is observed that $g(\xi) \approx f(\xi)$ for $\xi>\SI{0.3}{}$ for the symmetric  (Fig. \ref{fig:appC1}(a)) and asymmetric cases (Fig. \ref{fig:appC1}(b)). The left side of Eq. (\ref{eq:eq7}) (cf. insets of Fig. \ref{fig:appC1}) has values below \SI{0.4}{} and \SI{0.03}{} in the symmetric and asymmetric scenarios for the gate voltages used in each case. The latter indicates also that $g(\xi) \approx f(\xi)$ holds for all the devices and cases under study. 

These results validate Eq. (\ref{eq:Wd}) as equivalent to Eq. (\ref{eq:eq7}).

\end{document}